\documentclass[10pt]{article}
\usepackage{a4,amsmath,chicago,myequation}
\usepackage{calc,graphicx}

\def\IAUGC#1{\text{IAUGC}_{#1}}

\newcommand{\bld}[1]{\mathbf{#1}}
\newcommand{\bs}[1]{\boldsymbol{#1}}
\newcommand{\wh}[1]{\widehat{#1}}

\newcommand{\kep}{k_\text{ep}}
\newcommand{\vp}{v_p}
\newcommand{\ktrans}{K^\text{trans}}

\newcommand{\N}{\text{N}}

\newcommand{\IG}{\text{IG}}
\newcommand{\grad}{\text{\(\mathsurround=0pt{}^\circ\)}}
\newcommand{\distas}{\stackrel{\text{iid}}{\sim}}

\begin{document}

\title{A Bayesian Hierarchical Model for the Analysis of a
  Longitudinal Dynamic Contrast-Enhanced MRI Cancer Study} 

\author{Volker J.~Schmid$^\dagger$ \and Brandon Whitcher$^*$ \and Anwar
  R.~Padhani$^\ddagger$ \and N.~Jane Taylor$^\ddagger$ \and
  Guang-Zhong Yang$^\dagger$}
\date{\today}

\maketitle

\bigskip

\noindent $^*$ Corresponding Author:\\
\noindent Clinical Imaging Centre\\
\noindent GlaxoSmithKline\\
\noindent Hammersmith Hospital\\
\noindent Imperial College London\\
\noindent Du Cane Road\\
\noindent London ~~~ W12 0HS\\
\noindent United Kingdom\\
\noindent +44 20 8008 6052; +44 20 8008 6491 (fax)\\
\noindent {\em brandon.j.whitcher@gsk.com}

\bigskip

\noindent $^\dagger$ Institute of Biomedical Engineering\\
\noindent Imperial College London\\
\noindent London ~~~ SW7 2AZ\\
\noindent United Kingdom

\bigskip

\noindent $^\ddagger$ Paul Strickland Scanner Centre\\
\noindent Mt. Vernon Cancer Centre\\
\noindent Northwood ~~~ HA6 2RN\\
\noindent United Kingdom

\bigskip

\noindent
{\em Key words.} 

\medskip

\noindent
{\em Running head.} Bayesian Hierarchical Model for DCE-MRI in Oncology

\medskip

\noindent
{\em Approximate word count.} 5000

\newpage

\section*{Abstract}

Imaging in clinical oncology trials provides a wealth of information
that contributes to the drug development process, especially in early
phase studies.  This paper focuses on kinetic modeling in DCE-MRI,
inspired by mixed-effects models that are frequently used in the
analysis of clinical trials.  Instead of summarizing each scanning
session as a single kinetic parameter -- such as median $\ktrans$
across all voxels in the tumor ROI -- we propose to analyze all voxel
time courses from all scans and across all subjects simultaneously in
a single model.  The kinetic parameters from the usual non-linear
regression model are decomposed into unique components associated with
factors from the longitudinal study; e.g., treatment, patient and
voxel effects.  A Bayesian hierarchical model provides the framework
in order to construct a data model, a parameter model, as well as
prior distributions.  The posterior distribution of the kinetic
parameters is estimated using Markov chain Monte Carlo (MCMC) methods.
Hypothesis testing at the study level for an overall treatment effect
is straightforward and the patient- and voxel-level parameters capture
random effects that provide additional information at various levels
of resolution to allow a thorough evaluation of the clinical trial.
The proposed method is validated with a breast cancer study, where the
subjects were imaged before and after two cycles of chemotherapy,
demonstrating the clinical potential of this method to longitudinal
oncology studies.

\newpage

\section{Introduction}

Assessing the efficacy of cancer treatments using {\em in vivo}
imaging is shifting from qualitative techniques to quantitative
imaging methods that characterize biologically relevant properties of
tumor tissue.  The use of model-free or heuristic measures, such as
the initial area under the Gadolinium curve (IAUGC), or fully
quantitative measures, such as the kinetic parameters from a
compartmental model, are relatively well understood in the analysis of
dynamic contrast-enhanced magnetic resonance imaging (DCE-MRI)
\cite{padhani03,col-pad:ieee,tof-etal:estimating}.  Analysis of an
oncology imaging trial is usually achieved by applying statistical
summaries, such as the mean or median, to the parameters of interest
derived from tissue regions of interest (ROIs).  That is, enhancing
(tumor) voxels are identified from the DCE-MRI data for each scan
across all subjects and those voxels are represented by a single
parameter; e.g., $\ktrans$ from quantitative analysis and $\IAUGC{90}$
from a heuristic analysis.  Hypothesis testing, either parametric or
non-parametric, may then be applied to the derived statistics in order
to assess the effects of treatment.

Applying statistical summaries to the kinetic parameter maps from
DCE-MRI however discards a substantial amount of information contained
in the contrast agent concentration time curves (CTCs) at each voxel,
essentially abstracting thousands of observations in space and time to
a single number per scan per subject.  We believe that there is a
wealth of potential information by retaining the collection of CTCs
across all subjects and scans, acknowledging the fact that not all
CTCs are the same and not all patients are the same.

This paper proposes a Bayesian hierarchical model to analyze all tumor
CTCs across all patients and scans in a given study simultaneously
based on the concept of a mixed-effects model.  Mixed-effects models
are well established in the statistical community and have found
widespread applications in, for example, agriculture, economics,
geophysics and the analysis of clinical trials
\cite{pin-bat:book,bro-pre:mixed-models}.  Mixed-effects models extend
the concept of traditional linear or non-linear models by combining
both fixed effects and random effects in the same model.  More
generally, mixed-effects models are most often used to describe
relationships between the measured response and explanatory variables
in data that are grouped according to one or more factors.  Fixed
effects denote parameters that are associated with an entire
population and random effects denote parameters which are associated
with random samples from a population.  For example, the drug or
radiation therapy given in a trial is a fixed effect because there is
no randomness associated with it, whereas patients are inherently
random because they are sampled from the population of all patients
with a given disease.  By acknowledging the fact that some parameters
are associated with random samples from a population, we are able to
generalize the results from a mixed-effects model beyond the
collection of subjects used in the model fitting.

Bayesian methods are used in the construction and estimation of the
generalized additive model \cite{fahrmeir01a,lfahrmeir01b} associated
with each kinetic parameter in the model of the CTCs.  Similarly to
mixed-effects models in a maximum likelihood setting the variances
associated with the fixed effects are chosen to be constant, but the
variance terms associated with the random effects have prior
distributions.  This leads to a shrinkage estimation of the random
effects so that the random effects are pushed towards zero
\cite{tibshirani96}.  The fixed effect in the model explain as much
variance as possible, whereas the random effects capture variability
that cannot be explained by the fixed effects.

Formulation of a Bayesian hierarchical model is typically achieved in
three stages: the data model, the parameter model and the prior
parameters \cite{gel-etal:bayesian,wik:hierarchical}.  The data model
reflects our knowledge of the CTCs at the voxel level using the class
of compartmental models
\cite{ket:blood-tissue,tof-ker:measurement,lar-tof:measurement} with a
standard arterial input function (AIF) taken from the literature
\cite{wei-lan-mut:pharmacokinetics,fri-etal:measurement,buc-par:measuring}.
The process model describes how model parameters are generated from
underlying processes.  At this step we decompose the kinetic
parameters into treatment, patient and voxel effects.  In Bayesian
theory all parameters are regarded as random variables with
pre-specified (prior) distributions.  This includes the parameters for
the fixed as well as the random effects in the model.  Conjugate
distributions with appropriately selected values are used as priors on
all parameters in the model in order to limit their range during the
estimation procedure.  These choices also allow us to implement
efficient sampling methods, wherever possible, to reduce the
computational burden.

As an illustration of the parameter model let $\psi$ be a kinetic
parameter of interest and suppose the imaging study acquires a dynamic
MRI measurement sequence at two time points for each subject -- before
and after treatment.  The generalized additive model uses the natural
logarithm as the link function between the signal and effects that can
be identified from the study design.  For example, assume the imaging
study included two time points (pre- and post-treatment) so the
parameter $\psi$ may be decomposed via
\begin{equation}\label{eqn:psi}
  \ln\psi = \text{baseline} + \text{treatment} +
  \text{patient} + (\text{patient $\cdot$ treatment}) + \text{voxel}.
\end{equation}
With model fitting considering all effects related to the kinetic
parameter $\psi$, a curve that fits the observed data at a particular
voxel (associated with a specific scanning session and patient) can be
derived.  This is illustrated in Figure~\ref{fig:ctc-examples} where
pre- and post-treatment voxels have been selected from three subjects.
The solid line in each plot is the fitted curve to the CTC at a single
voxel including all effects in the model (Eq.~[\ref{eqn:psi}]).

One can construct different versions of $\psi$ that involve a subset
of terms in Eq.~[\ref{eqn:psi}].  For example, defining
$\hat\psi_b=\exp(\text{baseline})$ produces the estimated baseline
value of the kinetic parameter across the entire study, while ignoring
patient- and voxel-specific information and
$\hat\psi_t=\exp(\text{baseline}+\text{treatment})$ produces the
estimated treatment value of the kinetic parameter across the entire
study when patient- and voxel-specific information are ignored.  In
Figure~\ref{fig:ctc-examples}, the dotted line in each plot is the
fitted curve corresponding to the posterior median CTC for the whole
study pre-treatment (top row) and post-treatment (bottom row).

Relative changes between baseline and treatment are also available by
looking at the individual components in Eq.~[\ref{eqn:psi}]; e.g.,
$\hat\psi_{t^*}=\exp(\text{treatment})$ produces the percentage change
due to treatment relative to the baseline value $\hat\psi_b$.  Given
the mathematical properties of exponential functions, we know that
$\hat\psi_b \cdot \hat\psi_{t^*} = \hat\psi_t$, thus relating the
relative changes attributed to specific effects in the generalized
additive model to absolute values of the kinetic parameter.  Similar
manipulations may be performed to investigate patient- or
voxel-specific effects and their relative changes from baseline.  This
figure is described in greater detail in the results section.

The goal of model construction and estimation presented here is to
quantify the effect of drug treatment on disease -- in this case
breast cancer -- through quantitative summaries of tumor
microvasculature using DCE-MRI.  The model framework we have adopted
provides a unified treatment of imaging information at the study level
through simultaneous estimation of parameters at the voxel, patient
and treatment level, allowing a thorough interrogation of the results.

\section{Bayesian Hierarchical Model}

Bayesian methods rely on the specification of prior distributions
$p(\bs{\theta})$ that express our information about the unknown
parameters $\bs{\theta}$ before any measurements are obtained; i.e.,
our model assumptions.  To assess the model parameters after observing
the data, the posterior distribution $p(\bs{\theta}\,|\,\bld{Y})$ can
be computed, where $\bs{\theta}$ is the vector of all unknown
parameters and $\bld{Y}$ is the vector of observations.  The posterior
distribution of the parameter vector $\bs{\theta}$ is obtained by
applying Bayes' theorem
\begin{equation}
  p(\bs{\theta}\,|\,\bld{Y}) = \frac{p(\bs{\theta}) ~
    \ell(\bld{Y}\,|\,\bs{\theta})} {\int p(\bs{\theta}^*) ~
    \ell(\bld{Y}\,|\,\bs{\theta}^*)\,d\bs{\theta}^*},
\end{equation}
where $\ell(\bld{Y}\,|\,\bs{\theta})$ denotes the likelihood function
of $\bld{Y}$ and $p(\bs{\theta})$ the product of all {\em a priori}
probability distribution functions.  One can think of the posterior as
an update to the prior distribution, our beliefs, on $\bs{\theta}$
after measuring a process -- producing a mixture of previous knowledge
and experimental data.  Bayesian methods are inherently iterative,
since the posterior distribution can become our new prior distribution
and, be combined with new measurements of the data generating process
at a later date, to produce an updated posterior distribution.

The following sections introduce the key components in the Bayesian
hierarchical model: the data model, the parameter model and the prior
parameters.  Each stage of the model development has been tailored to
the analysis of a longitudinal cancer treatment study with two time
points.  Figure~\ref{fig:hierarchical} provides a schematic overview
of the proposed Bayesian Hierarchical Model (BHM).  The three model
stages are the rows and the columns represent the ``resolution'' of
the parameters.  For example, $\ktrans$ is decomposed into global
(study-wide), subject and voxel effects through the BHM where as $\vp$
is simply estimated for each voxel without further decomposition.  The
measurement error term is independent of the specific parameter model
and involves both prior and hyperprior distributions.  A standard
compartmental model is used to describe the concentration time curves
observed at each voxel.  A generalized additive model is proposed to
decompose the kinetic parameters into factors that are relevant to the
design of the longitudinal study.  Finally, the prior distributions,
including necessary hyperparameters, are specified on all factors of
the parameter model.  These prior distributions are flat in most
cases, reflecting a lack of knowledge concerning the parameter, but
also incorporate biological knowledge, such as a transfer rate must be
non-negative, or statistical knowledge, for example a variance must be
non-negative.

\subsection{Data Model}

A hierarchical Bayesian framework is used to model the contrast agent
concentration time curve (CTC) of all voxels \cite{gelman03}.  Let
$\bld{Y}=[Y(t_1), Y(t_2), \ldots, Y(t_T)]^\mathsf{T}$ denote the CTC
associated with a single voxel observed at $T$ time points determined
by the image acquisition protocol.  The CTC is assumed to follow a
standard compartment model \cite{buc-par:measuring}
\begin{equation}\label{eqn:compartmental-model}
  C_t(t) = v_p C_p(t) + C_p(t) \otimes \ktrans \exp(-t\,\kep),
\end{equation}
where $\otimes$ denotes the convolution operator, $\ktrans$ represents
the transfer rate from plasma to extracellular extravascular space
(EES) per minute, $\kep$ the rate constant between EES and blood
plasma per minute and $v_p$ the vascular space fraction.  The choice
of model for the CTC depends on the scientific goals of the study.
Replacing Eq.~[\ref{eqn:compartmental-model}] with a more or less
complicated model is straightforward in this model-building framework.
The observed vector $\bld{Y}$ may therefore be thought of as noisy
observations of the true contrast agent concentration $C_t(t)$ given
by a draw from a multivariate Normal distribution
\begin{equation}\label{eqn:data-model}
  \bld{Y} \sim \N_T\!\left( \bld{C}_t, \sigma^2I_T \right),
\end{equation}
where the notation $Y\sim{\N(\mu,\sigma^2)}$ means that the random
variable $Y$ is drawn from the Normal distribution with parameters
$\mu$ and $\sigma^2$.

We assume a common arterial input function (AIF), taken from the
literature for all patients in the study, and we follow the work of
Tofts and Kermode \cite{tof-ker:measurement} by using a bi-exponential
function
\begin{equation}\label{eqn:AIF}
  C_p(t) = D [ a_1 \exp(m_1t) + a_2 \exp(m_2t) ],
\end{equation}
where $a_1=24.0~\text{kg/l}$, $a_2=6.20~\text{kg/l}$,
$m_1=3.00~\text{min}^{-1}$ and $m_2=0.016~\text{min}^{-1}$ are
inspired by the work of Fritz-Hansen {\em et al.}
\cite{fri-etal:measurement}.

A Bayesian implementation of the compartmental model above was
proposed in Schmid {\em et al.} \cite{schmid06}.  Since the Bayesian
model framework does not depend on any optimization procedure, it will
produce valid parameter estimates when estimation via nonlinear
regression fails to converge.  Samples from the posterior distribution
are built up during the model fitting procedure for each parameter.
Hence, the posterior distribution may be used to obtain additional
information on the accuracy and precision of the estimates.  For
example, the standard error of the posterior is the estimation error.
Statistics of interest may be derived from the posterior distribution
(e.g., mean, median, quantiles, etc.~) so that not only point
estimates but also confidence intervals are readily available.

\subsection{Parameter Model}
 
The pharmacokinetic (PK) parameters from the data model are estimated
at every tumor voxel across all subjects and scans.  We assume {\em a
priori} that the distribution of the random variables $\ktrans$ and
$\kep$ in the tumor are patient-specific and are changed by treatment
in a similar way.  Therefore a generalized additive model is used
where the log-transformed kinetic parameters $\ln(\ktrans)$ and
$\ln(\kep)$ are expressed as a linear combination of fixed- and
random-effects associated with identifiable factors in the study.  In
addition to mathematical convenience, the log transform is also
appealing in this context because individual terms in the additive
model may be interpreted as percentage change from baseline.  We
assume that the distribution of the vascular fraction $v_p$ will not
be changed by the treatment, however single $v_p$ values will be
changed.  Let $i=1,\ldots,I$ denote the scans acquired and let
$j=1,\ldots,J$ denote the patients, so that $n_{ij}$ denotes the
number of tumor voxels for patient $j$ at scan $i$, measured at $T$
time points.  The transfer rate constants in
Eq.~[\ref{eqn:compartmental-model}] are assumed to be non-negative and
estimated in log-transformed space.  That is, let
$\psi_1=\ln(\ktrans$), $\psi_2=\ln(\kep)$ and
$\bs{\psi}=[\ln(\ktrans),\ln(\kep)]^\mathsf{T}$.

\begin{figure}[p]
  \begin{center}
    \includegraphics[width=.75\textwidth]{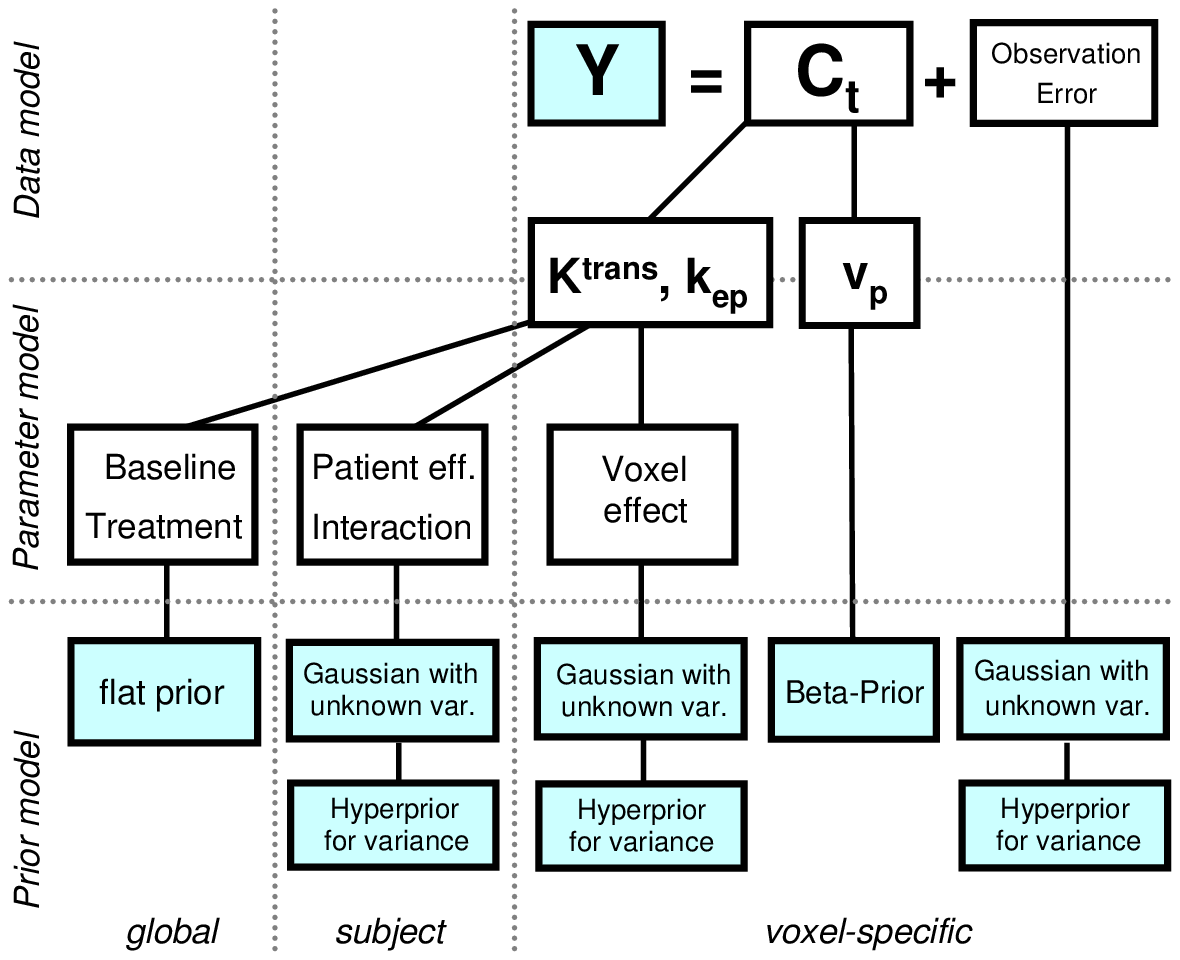}
  \end{center}
  \caption{}
  \label{fig:hierarchical}
\end{figure}

The factor of interest when measuring a change in the kinetic
parameters is the treatment effect, or the difference between
scan~$i=1$ and scan~$i=2$ when only pre- and post-treatment images are
acquired.  We acknowledge the fact that substantial variability exists
across patients in the study and between the voxels in each region of
interest (ROI) that describes the enhancing region in the acquisition.
Hence, the model for $\ln(\ktrans)$ is given by
\begin{equation}
  \psi_{ijk1} = [ 1 ~ x_i ] \left[ \begin{array}{cc}
      \alpha_1\\
      \beta_1
      \end{array}\right] + [ 1 ~ x_i ] \left[ \begin{array}{cc}
      \gamma_{j1}\\
      \delta_{j1}
      \end{array}\right] + \epsilon_{ijk1}, 
  \quad \text{for all $i,j,k$}
\end{equation}
where
\begin{equation}
  x_i = \left\{ \begin{array}{ll}
    1 & \text{scan $i=2$};\\
    0 & \text{otherwise}.
    \end{array} \right.
\end{equation}
The parameter $\alpha_1$ is the value of $\ln(\ktrans)$ associated
with the baseline scan and $\beta_1$ is the treatment effect (since it
is only associated with the post-treatment acquisition).  These
parameters are regarded as fixed effects (the global column of
Figure~\ref{fig:hierarchical}), and thus do not vary between patients
in the study.  In the Bayesian framework, a marginal posterior
distribution will be available for each parameter.  The parameter
$\gamma_{j1}$ is the effect of patient~$j$ on $\ln(\ktrans)$ and
$\delta_{j1}$ is the interaction between patient~$j$ and treatment.
These parameters are random effects since each patient is assumed to
be drawn from a larger population of patients suffering from this
condition (the subject column of Figure~\ref{fig:hierarchical}).
Finally, the parameter $\epsilon_{ijk1}$ is the random effect of
voxel~$k$ in scan~$i$ of patient~$j$ on $\ln(\ktrans)$.  The voxel
effect acknowledges the fact that each voxel in the tumor volume is
drawn from a distribution that describes the ideal tumor voxel (the
voxel column of Figure~\ref{fig:hierarchical}).  The combination of
fixed and random effects in a single model is commonly referred to as
a mixed-effects model \cite{fahrmeir01a}.

Using matrix notation, we can combine the generalized additive model
across both kinetic parameters, $\ln\ktrans$ and $\ln\kep$, such that
\begin{equation}\label{eqn:parameter-model}
  \bs{\psi}_{ijkl} = \bld{Z}_i \left[ \begin{array}{cc}
      \bs{\phi}\\
      \bs{\theta}_j
      \end{array}\right] + \bs{\epsilon}_{ijkl}
\end{equation}
\begin{equation}
  \bld{X}_i = \left[ \begin{array}{cccc}
      1 & x_i & 0 & 0\\
      0 & 0 & 1 & x_i
      \end{array}\right]; \quad
  \bld{Z}_i = [\bld{X}_i ~ \bld{X}_i]; \quad
  \bs{\phi}_l = \left[ \begin{array}{cc}
      \alpha_1\\
      \beta_1\\
      \alpha_2\\
      \beta_2
      \end{array}\right]; \quad
  \bs{\theta}_{jl} = \left[ \begin{array}{cc}
      \gamma_{j1}\\
      \delta_{j1}\\
      \gamma_{j2}\\
      \delta_{j2}
      \end{array}\right]; \quad
  \bs{\epsilon}_{ijkl} = \left[ \begin{array}{cc}
      \epsilon_{ijk1}\\
      \epsilon_{ijk2}
      \end{array}\right]
\end{equation}
The scan-specific covariates in the model are captured in $\bld{Z}_i$,
the fixed effects are in $\bs{\phi}_l$, the patient-specific random
effects are in $\bs{\theta}_{jl}$ and the voxel-specific random
effects are in $\bs{\epsilon}_{ijkl}$.  The model formulation in
Eq.~[\ref{eqn:parameter-model}] can be adapted in order to incorporate
a greater number of scans in a longitudinal study.

\subsection{Prior Models}

In the Bayesian framework prior information with unknown variance is
used to model the random effects.  We use vector notation to denote
the patient-specific random effects such that
\begin{equation}
  \bs{\gamma} = \left[ \begin{array}{c}
      \gamma_{11}\\ \gamma_{21}\\ \vdots\\ \gamma_{J1}\\ \gamma_{12}\\
      \vdots\\ \gamma_{J2}
      \end{array}\right] \quad \text{and} \quad
  \bs{\delta} = \left[ \begin{array}{c}
      \delta_{11}\\ \delta_{21}\\ \vdots\\ \delta_{J1}\\ \delta_{12}\\
      \vdots\\ \delta_{J2}
      \end{array}\right],
\end{equation}
where we have dropped the kinetic parameter subscript to simplify the
notation.  We draw from multivariate Gaussian distributions to
characterize the prior distribution of the unknown variances for the
patient-specific random effects, i.e.,
\begin{eqnarray}
  \bs{\gamma} &\sim& \N_{2J}\!\left( \bld{0},
  \text{diag}\!\left(\bs{\tau}^2_\gamma\right) \right),\\
  \bs{\delta} &\sim& \N_{2J}\!\left( \bld{0},
  \text{diag}\!\left(\bs{\tau}^2_\delta\right) \right),
\end{eqnarray}
where $\bs{\tau}^2_\gamma$ and $\bs{\tau}^2_\delta$ are vectors of the
same length and indexed as $\bs{\gamma}$ and $\bs{\delta}$,
respectively.  The voxel-specific random-effect vectors are given
unique prior distributions by scan, patient and parameter, so that
each vector is given by $\bs{\epsilon}_{ijl}=[\epsilon_{ij1l},
\epsilon_{ij2l}, \cdots, \epsilon_{ijn_{ij}l}]^\mathsf{T}$ and it is
drawn from a multivariate Gaussian distribution via
\begin{equation}
  \bs{\epsilon}_{ijl} \sim \N_{n_{ij}}\!\left( \bld{0},
  \tau_{\epsilon;ijl}^2 I_{n_{ij}}\right)
\end{equation}
where $n_{ij}$ is the number of voxels in the region of interest of
scan~$i$ of patient~$j$, and $\bs{\tau}_{\epsilon;ijl}^2$ is the
unknown variance associated with scan~$i$, patient~$j$ and
parameter~$l$.  Since the variances are unknown parameters, they must
have their own prior distributions which are given by
\begin{eqnarray}
  \bs{\tau}^2_\gamma & \distas & \IG\!\left( 1, 1 \right),\\ 
  \bs{\tau}^2_\delta & \distas & \IG\!\left( 1, 1 \right),\\ 
  \bs{\tau}^2_\epsilon & \distas & \IG\!\left( 1, 10^{-5} \right),
\end{eqnarray}
where $\IG(a,b)$ denotes the Inverse Gamma distribution
\cite{joh-kot-bal:univariate1}, allowing only non-negative values.
The inverse Gamma distribution is a conjugate prior for the Normal
distribution.  For the fixed effects we use flat priors; i.e., the
prior distribution does not contain any relevant information, such
that
\begin{equation}
  p(\alpha_l) = p(\beta_l) = \text{constant} \quad \text{for $l=1,2$}.
\end{equation}
The prior distributions on the coefficients in the generalized
additive model are chosen so that as much variance in the data is
explained by the fixed effects $\alpha$ and $\beta$ -- as no prior
information is used for those parameters.  Variability which cannot be
explained by the fixed effects will be covered by the random effects
$\gamma$ and $\delta$.  For these parameters an additional prior
distribution (hyperprior) on the variance of the parameters is
defined, which leads to a shrinkage of those effects, that is the
parameters are pushed towards zero and therefore do not cover variance
explained by the fixed effects.  Any additional variance is
explained by the voxel effects. 

 For the vascular space fraction we
impose a relatively flat prior
\begin{equation}
  v_{p;ijk} \distas \text{B}(1,19), \quad \text{for all $i,j,k$},
\end{equation}
where $\text{B}(a,b)$ denotes the Beta distribution
\cite{joh-kot-bal:univariate2}, so that the {\em a priori} expected
value of $v_p$ is 0.05.  The Bayesian hierarchical model is complete
by specifying a prior distribution for the variance of the
observational error in Eq.~[\ref{eqn:data-model}],
with one variance parameter per scan per patient,
\begin{equation}
  \sigma^2_{ij} \distas \IG(1,10^{-2}) \quad \text{for all $i,j$}.
\end{equation}

\section{Materials and Methods}

\subsection{Data acquisition}

The first twelve patients from a previously reported breast cancer
study are included in the analysis \cite{ahsee04,schmid06}.  Data were
provided by the Paul Strickland Scanner Centre (PSSC) at Mount Vernon
Hospital, Northwood, UK.  Each patient underwent a DCE-MRI study
before and after two cycles of chemotherapy (5-fluorouracil,
epirubicin and cyclophosphamide).  Six of these patients were
identified as pathological responders after receiving six cycles of
chemotherapy, the others were non-responders.

For the calculation of $T_1$ values, we used a two-point measurement
with calibration curves as described in \cite{parker97,darcy06}.  The
$T_1$ values are computed as ratio of a $T_1$-weighted fast low-angle
shot (FLASH) image and a proton density weighted (PDw) FLASH image.
The imaging parameters of the $T_1$-weighted FLASH images were
$\text{TR}=11~\text{ms}$, $\text{TE}=4.7~\text{ms}$, $\alpha=35\grad$,
the parameters of the proton density-weighted image were
$\text{TR}=350~\text{ms}$, $\text{TE}=4.7~\text{ms}$, $\alpha=6\grad$.
Field of view was the same for all scans, $260 \times 260 \times
8~\text{mm}$ per slice, so voxel dimensions were $1.016 \times 1.016
\times 8~\text{mm}$.  A scan consists of three sequential slices of
$256 \times 256$ voxels and one slice placed in the contra lateral
breast as control, which we do not use for our analysis.  A total of
$40$ to $50$ acquisitions were acquired, with one acquisition each
11.9 seconds.  A dose of $D=0.1~\text{mmol}$ per kg body weight of
Gd-DTPA was injected after the fourth scan using a power injector with
$4~\text{ml/s}$ with a $20~\text{ml}$ saline flush also at
$4~\text{ml/s}$.  The first four scans, before contrast, were used to
compute $T_{10}$ as the average of the $T_1$ values of these images.
Data from this study were acquired in accordance with the
recommendation given by \cite{leach05}.  Informed consent was obtained
from all patients.

Regions of interest (ROIs) were drawn manually by an expert
radiologist on a scan-by-scan basis using anatomical images and 
subtraction images from the
dynamic data to define tumor voxels in pre and post treatment scans.

\subsection{Parameter Estimation via MCMC}

The joint posterior distribution of all parameters was assessed using
Markov chain Monte Carlo (MCMC) {\cite{gilks96}}.  After a initial
burn-in phase of 10,000 iterations, another 100,000 iterations were
computed.  To ensure independent samples only each 100th sample was
used, giving us a total of 1000 samples to describe the posterior
distribution.  The global parameters $\bs{\phi}$ and patient-specific
$\bs{\theta}_j$ were drawn {\em en bloc} in Gaussian Gibbs steps
\cite{liu94}, and hyperparameters were drawn in independent Gamma
Gibbs steps; technical details can be found in Appendix~\ref{app:fc}.
Metropolis-Hastings steps with random walk proposals were necessary
for the voxel-specific random effects and vascular space fraction.
The algorithm was tuned to an acceptance rate of 30-50\%
{\cite{schmid05a}}.  Summary statistics were computed from the
samples of the posterior distribution to provide point estimates of
the parameters.  Empirical standard errors, along with sample
quantiles, were used to characterize the precision of the parameter
estimates.

\section{Results}

All parameter estimates are derived from the posterior distribution
using Bayes theorem.  Hence, a sampling distribution for each
parameter value has been built up from which we can produce a point
estimate via the median of the sample and also credible intervals
(Bayesian confidence intervals) by using the quantiles from their
sampling distributions.

\begin{figure}[p]
  \begin{center}
    \includegraphics*[width=\textwidth]{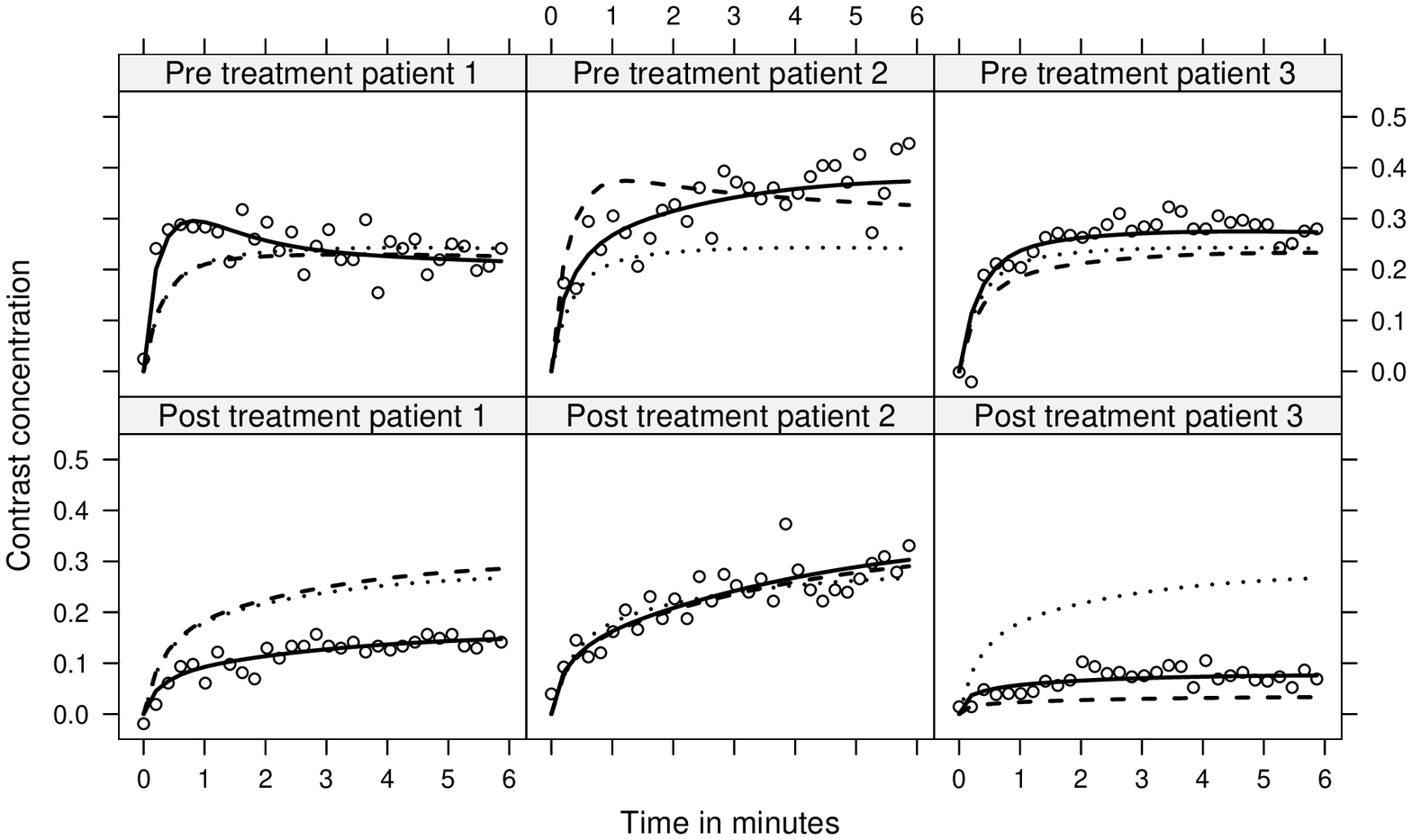}
  \end{center}
  \caption{}
  \label{fig:ctc-examples}
\end{figure}

How the individual parameters from the generalized additive model
coalesce to fit the observed contrast agent concentration time curve
is illustrated, at the voxel level, in Figure~\ref{fig:ctc-examples}.
The observed CTCs for two voxels from three subjects, one voxel at
baseline and one voxel after treatment, are plotted along with three
fitted curves.  The best estimate from the Bayesian hierarchical model
at a specific voxel is provided by the solid lines in each plot.  That
is, all parameters from the generalized additive model
(Eq.~[\ref{eqn:parameter-model}]) are used in the parameter model
in order to fit the data model.  These curves are very similar to, but
not exactly the same as, model fits from the standard non-linear
regression method used in the quantitative analysis of DCE-MRI data
\cite{schmid06}.  Removing the voxel-specific term from the model
produces a fitted curve that is associated with patient and treatment
effects, but not the specific voxel, and are plotted as dashed lines
in Figure~\ref{fig:ctc-examples}.  Given the presence of inter-voxel
heterogeneity in the tumor ROI, the dashed lines may or may not fit
the observed data at a given voxel very well but they do represent the
best (in the sense of a posterior median) fit to all voxels in the
tumor ROI for a given patient at a single scan time point.  Going back
one more level in the generalized additive model and removing the
patient effect leaves a fitted curve associated with the baseline and
post-treatment scans (i.e., two curves that summarize the overall
treatment effect) given by the dotted lines.  The top row of
Figure~\ref{fig:ctc-examples} contains voxels from three subjects
before treatment so the dotted lines are identical and represent the
best (in the sense of a posterior median) fit to all pre-treatment
voxels across all subjects.  The bottom row contains voxels from the
same subjects after treatment and the dotted line is the best fit to
all post-treatment voxels.

\begin{figure}[p]
  \begin{center}
    \includegraphics*[width=.8\textwidth]{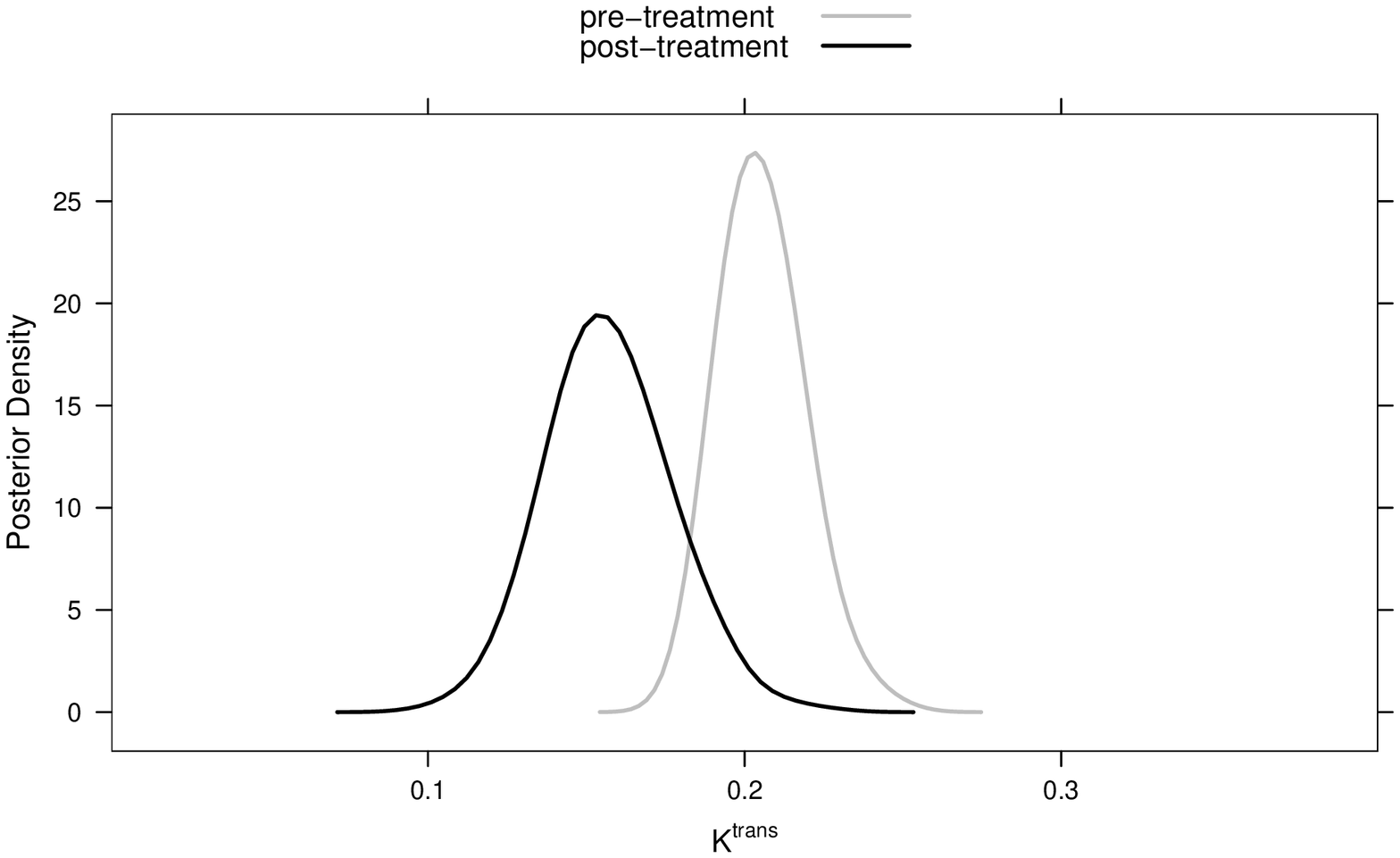}
  \end{center}
  \caption{}
  \label{fig:treatment-effects}
\end{figure}

Figure~\ref{fig:treatment-effects} shows the posterior distributions
of pre-treatment (baseline) $\ktrans$ and post-treatment $\ktrans$.
That is, the posterior samples were transformed via $\exp(\alpha_1)$
and $\exp(\alpha_1+\beta_1)$, respectively.  For ease of comparison
between the two posterior distributions a smoothed version of each
histogram, known as a kernel density estimate, is displayed
{\cite{sil:density}}.  The posterior median of $\ktrans$ at baseline
is 0.205, and the posterior median of $\ktrans$ after treatment is
0.156.  Credible intervals for $\ktrans$, that cover 95\% of the
posterior distribution, are $[0.186,0.234]$ at baseline and
$[0.121,0.198]$ after treatment.  A credible interval is a posterior
probability interval.  That is, the true value of $\ktrans$ lies in
the interval $[0.186,0.234]$ with posterior probability 0.95 at
baseline and in $[0.121,0.198]$ with posterior probability 0.95 after
treatment.

The density estimates in Figure~\ref{fig:treatment-effects} are
unimodal and indicate an overall decrease in $\ktrans$ after
treatment.  In order to test for a treatment effect on $\ktrans$,
specifically a reduction in $\ktrans$ in the second acquisition
compared to the first, we construct the hypothesis 
\begin{equation}\label{eqn:H0}
H_0 : \beta_1 > 0 \quad \text{versus} \quad H_1 : \beta_1 \leq 0,
\end{equation}
using the treatment effect from the parameter model
(Eq.~[\ref{eqn:parameter-model}]) and calculate the posterior
probability of $\beta_1$ exceeding zero.  From the results of the MCMC
simulation, the null hypothesis (Eq.~[\ref{eqn:H0}]) is rejected with
$p=0.001$.

When introducing the generalized additive model previously the fact
that the parameter $\ktrans$ and the covariates are linked through a
logarithmic transform leads to the interpretation of individual
covariates in the parameter model as percentage changes from baseline
instead of absolute changes.  For the treatment effect this translates
into a $100\%\cdot|0.7659-1|=23.3$\% median reduction in $\ktrans$
from baseline, where the sign determines whether the change is
associated with an increase or decrease.

\begin{figure}[p]
  \begin{center}
    \includegraphics*[width=\textwidth]{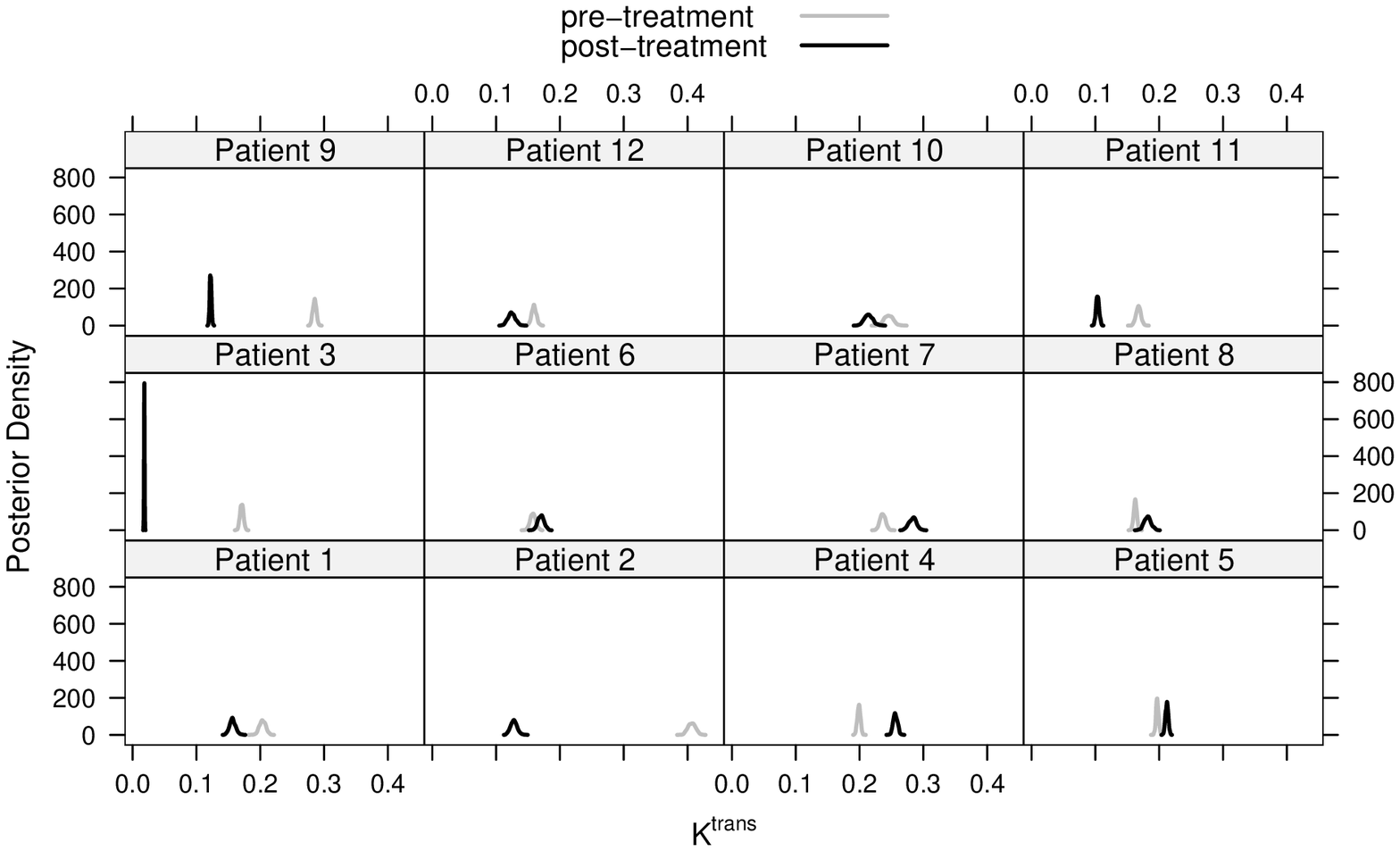}
  \end{center}
  \caption{}
  \label{fig:patient-effects}
\end{figure}

Figure~\ref{fig:patient-effects} shows the patient-specific posterior
distributions for pre-treatment $\ktrans$, given by
$\exp(\alpha_1+\gamma_{j1})$ for $j=1,\ldots,12$, and post-treatment
$\ktrans$, given by $\exp(\alpha_1+\beta_1+\gamma_{j1}+\delta_{j1})$
for $j=1,\ldots,12$.  The clinical responders are grouped in the first
two columns of Figure~\ref{fig:patient-effects} and the clinical
non-responders are in the third and fourth columns.  The same range
for $x$-axis $[0,0.45]$ was used in all plots of $\ktrans$ for
comparison.  In general the decrease in $\ktrans$ observed in the
clinical responders is greater than the clinical non-responders, but
this is not absolute.  For example, patient~12 shows only a small
decrease in $\ktrans$ post-treatment and patient~6 shows an increase
in $\ktrans$ after treatment, but both are clinical responders after
additional chemotherapy.  The interpretation of the treatment effect
as a percentage change from baseline helps to quantify the results in
Figure~\ref{fig:patient-effects}.  The median percentage change in
$\ktrans$ for subject~$j$ is obtained via
$100\%\cdot|\exp(\hat\beta_1+\hat\delta_{j1})-1|$, where the sign
determines whether an increase or decrease occurred.  For example,
patient~1 (pathological responder) experienced a
$100\%\cdot|0.7684-1|=23.2$\% median reduction in $\ktrans$ which is
very similar to the overall treatment effect.  This is definitely not
the norm as patient~9 experienced a $100\%\cdot|0.4285-1|=57.2$\%
median reduction in $\ktrans$ and patient~6 experienced a
$100\%\cdot|1.0817-1|=8.17$\% median {\em increase} in $\ktrans$, both
were pathological responders.

\begin{figure}[p]
  \begin{center}
    \includegraphics*[width=\textwidth]{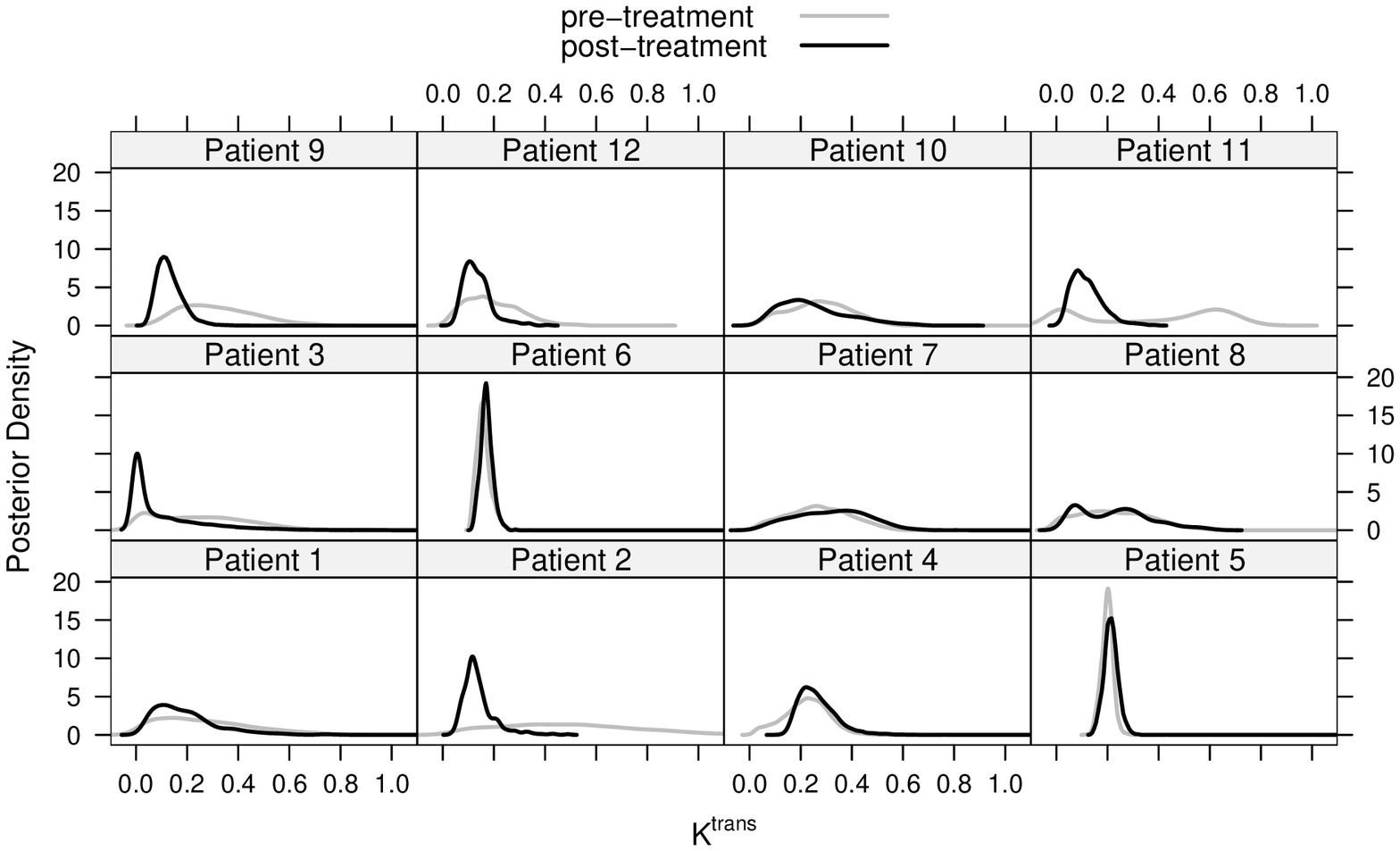}
  \end{center}
  \caption{}
  \label{fig:voxel-effects}
\end{figure}

Figure~\ref{fig:voxel-effects} shows the voxel-specific median
posterior for pre- and post-treatment $\ktrans$.  The clinical
responders are grouped in the first two columns and the clinical
non-responders are in the third and fourth columns (identical to
Fig.~\ref{fig:patient-effects}).  The range for the $x$-axis was
restricted to $[0,1]$ in all plots for comparison.  Given the number
of samples from the posterior distribution across all voxels, the
median value of $\exp(\alpha_1+\gamma_{j1}+\epsilon_{1jk1})$ for
$j=1,\ldots,12;k=1,\ldots,n_{1j}$ and
$\exp(\alpha_1+\beta_1+\gamma_{j1}+\delta_{j1}+\epsilon_{2jk1})$ for
$j=1,\ldots,12;k=1,\ldots,n_{2j}$ across the 1000 samples, for each
voxel $k$, was computed to summarize the voxel effect.  The resulting
histograms for the voxel effect have been summarized by a kernel
density estimate.  Most voxel-level distributions of median~$\ktrans$
show a substantial change in shape after treatment, although this is
more apparent in the responders compared to the non-responders.  It is
interesting to note the extent of changes in the shape of these
distributions between the different subjects.  For example, patient~11
is characterized by a tumor with two distinct modes in estimated
$\ktrans$ at baseline and a single mode after treatment.  Even more
interesting is the fact that the post-treatment distribution of
$\ktrans$ is in between the two modes at baseline.  The distributions
of median~$\ktrans$ for patient~12 show the reverse effect, albeit
much more subtle than patient~11, where the post-treatment
distribution of median~$\ktrans$ appears to be bimodal but still spans
a similar range of values.

\section{Discussions and conclusions}

Information is obtained at multiple levels during an imaging study in
the clinical trial setting.  The main scientific question of interest
is usually, was there a treatment effect?  This key hypothesis test
drives study design by influencing critical experimental design
parameters such as power and sample size.  However, information at
other levels, such as the patient or voxel level, can provide insight
into much more subtle features concerning patients, tumors and the
treatment effect.  Patient variability with application to predicting
clinical response and tumor heterogeneity, as measured by voxel-wise
properties of the pharmacokinetic model, are just two examples of
so-called secondary endpoints.

The Bayesian hierarchical model presented here was developed to test
the hypothesis of a treatment effect for an imaging study while
acknowledging known sources of uncertainty; e.g., patients and voxels.
This is similar to the approach taken in standard analysis methods for
clinical trials where fixed and random effects are identified in the
model.  The specification of fixed and random effects allows the
results from the study to be applicable beyond the specific patient
population recruited for this specific study. 

A standard analysis was performed on the ROIs and the median $\ktrans$
values have been summarized in Table~\ref{tab:standard-analysis}.  A
non-parametric test (a one-sided Wilcoxon signed rank test) was
performed to test that the difference between the median values was
greater than zero; i.e., the treatment did not reduce $\ktrans$ across
all subjects.  The null hypothesis was rejected at a borderline
significance level $(p=0.055)$.  Given the small sample size, $N_1=6$
responders and $N_2=6$ non-responders, this is an impressive result
and there is obviously a reasonable difference in $\ktrans$ between
the two groups.

\begin{figure}[p]
  \begin{center}
    \includegraphics*[width=\textwidth]{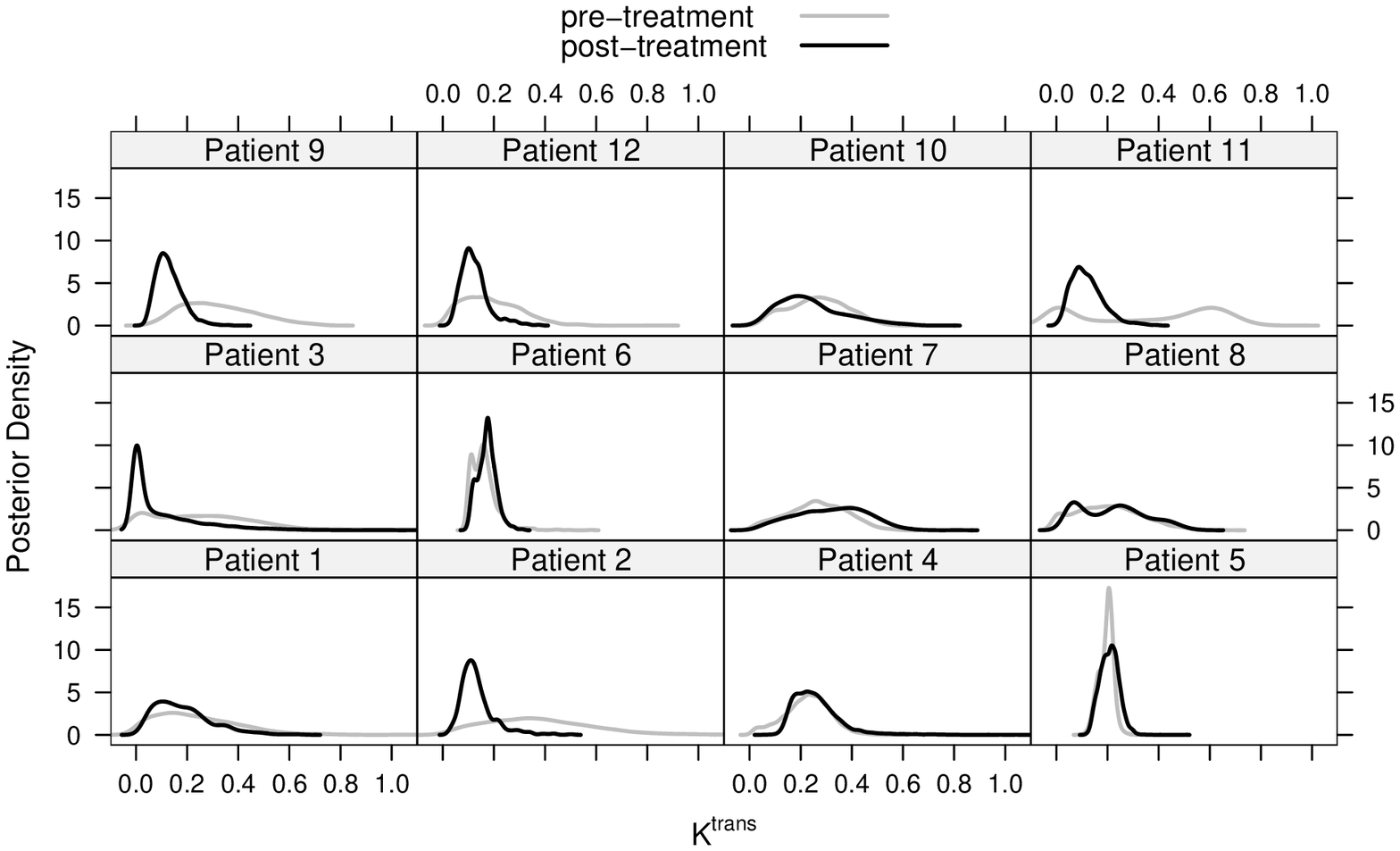}
  \end{center}
  \caption{}
  \label{fig:standard-analysis}
\end{figure}

Figure~\ref{fig:standard-analysis} shows the kernel density estimates
of $\ktrans$ for each ROI, before and after treatment, using a
voxel-wise non-linear regression analysis.  That is, the compartmental
model in Eq.~[\ref{eqn:compartmental-model}] was fit to each voxel
independently using the Levenberg-Marquardt optimization procedure.
The empirical distributions observed for each patient are extremely
similar to those obtained in the BHM.  This is to be expected given
the relatively flat priors that were imposed on the kinetic parameters
\cite{schmid06}.

While the voxel-wise results from the Bayesian and regression methods
are very similar, and thus provide a check on the consistency of the
Bayesian model fitting procedure, the advantages of the Bayesian
hierarchical model are clear through the coefficients from the
generalized additive model (Eq.~[\ref{eqn:parameter-model}]).  The
regression analysis can only summarize the study through
Table~\ref{tab:standard-analysis}, but the BHM allows one to isolate
and interrogate specific effects, at the study or patient or voxel
level, through the generalized additive model.  Examples of such
interrogations have been presented here in
Figures~\ref{fig:treatment-effects} and~\ref{fig:patient-effects}, but
the possibilities for such model summaries are only limited by the
construction of the parameter model.  

Bayesian models rely on {\em a priori} beliefs about the model and
parameters, expressed as prior distributions.  In general, a flat
prior provides similar information to a maximum likelihood approach,
and hence similar results.  However, in the Bayesian hierarchical
model proposed here the choice of prior distribution is critical in
specifying the model.  We used flat priors on the baseline $\alpha$
and on the treatment effect $\beta$, and thus the approach is similar
to a non-Bayesian or frequentist approach.  For the patient specific
effects $\gamma$ and $\delta$ we used Gaussian priors with unknown
variances; this is also known as shrinkage prior, as it shifts the
parameters towards zero.  Hence the patient-specific effects only pick
up the deviation from baseline and treatment effect.  The voxel effect
was also given a shrinkage prior with a more informative hyperprior
distribution on the variance, hence it only picks up variability after
modelling the baseline, treatment and patient-specific effects.

In this paper a generalized additive model was constructed for the
kinetic parameters ($\ktrans$ and $\kep$) in a compartmental model.
This model incorporated two scanning sessions, and all subjects, to
asses the effect of treatment.  The modeling framework is easily
extended to handle additional covariates or scanning sessions.  For
example, a dose-ranging study design could be incorporated into the
additive model where the treatment effect can be expressed as a
function of the dose.  Additional scans over time would enable the
assessment of temporal dependence on treatment and provide information
about the reliability of the data by potentially reducing the amount
of uncertainty in the parameter estimates.

Another possible extension of this model would be to include the
spatial information of adjacent voxels.  In the current implementation
of the Bayesian hierarchical model all voxels from one region of
interest (tumor) were treated as spatially independent.  Since voxel
borders are arbitrary and do not represent physiological boundaries
between different tissue types, it is likely that neighboring voxels
share similar perfusion characteristics.  This fact has been taken
advantage of in the context of Bayesian modeling of individual scans
from a DCE-MRI study {\cite{schmid06}}.  The inclusion of a
neighborhood structure in the modeling process would reduce the
uncertainty in estimation and provide more reliable estimates of the
kinetic parameters.

\section{Acknowledgements}

Support for VJ~Schmid was financed through a research grant from
GlaxoSmithKline.

\appendix

\section{Appendix}

\subsection{Full conditional distributions}
\label{app:fc}

In each iteration of the MCMC (Markov chain Monte Carlo) algorithm, a
random sample of the marginal posterior distribution for all
parameters is drawn.  This is performed by drawing from the
conditional posterior distribution of one or more parameters given all
other parameters and the data.  Hence, the full conditional
distributions must be computed.  The full conditional is denoted by
$\theta\,|\,\cdot$, where $\theta$ is the parameter and $\cdot$
denotes all other parameters and the data.  If the full conditional
takes the from of a standard distribution, one can sample directly
form this distribution; this is known as the Gibbs sampler
{\cite{gilks96}}.  If the full conditional is not a standard
distribution, then a Metropolis-Hastings sampler must be constructed.

In the proposed Bayesian hierarchical model all full conditionals are
from standard distributions due to the use of conjugate prior
distributions, except for the voxel effect and $v_p$.  Let
$\bs{\xi}_l=(\alpha_l,\beta_l,\bs{\gamma}_l,\bs{\delta}_l)$ denote the
vector of length~$P=I(J+1)$ associated with all parameters in the
generalized additive model, except the voxel effect, for a specific
kinetic parameter. The full conditional of $\bs{\xi}_l$ is a
multivariate Normal distribution given by
\begin{eqnarray*}
  \bs{\xi}_l \,|\, \cdot & \sim & \N_P \left(\bld{V}^{-1}
  \bld{m}, \bld{V}^{-1}\right),\\
  \bld{m} &=& [m_1,\ldots,m_P]^\mathsf{T},\\
  m_p &=& \sum_{i=1}^2\sum_{j=1}^J \left( \tau_{\epsilon;ij}
  \sum_{k=1}^{n_{ij}} w_{ijp}\psi_{ijkl} \right) \quad
  \text{for $p=1,\ldots,P$},\\
  \bld{V} &=& \bld{W}^\mathsf{T} \Lambda \bld{W} + \text{diag}\,(0, 0,
  \tau_{\gamma;1l}, \ldots, \tau_{\gamma;Jl}, \tau_{\delta;1l},
  \ldots, \tau_{\delta;Jl}),
\end{eqnarray*}
where $\bld{W}$ is a $I(J+1) \times P$ matrix indicating which
covariate should be included in the parameter model
(Eq.~[\ref{eqn:parameter-model}]) and $\Lambda$ is a diagonal matrix
with elements $n_{ij}\tau_{\epsilon;ij}$.  The vector $\bs{\xi}_l$ is
drawn in one block from a multivariate Normal distribution with an
efficient block-sampling algorithm \cite{rue01}.

The full conditional distribution of the voxel effect
$\epsilon_{ijkl}$ is a non-standard distribution.  For computational
reasons it is more convenient to sample from $\psi_{ijkl}$ rather than
from $\epsilon_{ijkl}$. where the full conditional distribution of
$\psi_{ijkl}$ is given by
\begin{equation}
  p(\psi_{ijkl}\,|\,\cdot) \propto \exp\left( -\frac{1}{2}
  \tau_{\epsilon;ijk} \psi_{ijkl}^2 - \frac{1}{2\sigma^2_{jk}}
  \left(Y_{ijkl} - \wh{Y}_{ijkl}\right)^2 \right).
\end{equation} 
Note, $\wh{Y}_{ijkl}$ is the estimated contrast agent concentration
curve given by the estimated model parameters in $\hat\psi_{ijkl}$.
Draws from this distribution are obtained using a Metropolis-Hastings
step.

The full conditionals of all variance parameters are inverse Gamma
distributions, which are given by 
\begin{eqnarray}
  \tau^2_{\gamma} \,|\, \cdot & \distas & \IG\left( 1.5, 1 +
  \gamma_{jl}^2 \right),\\
  \tau^2_{\delta} \,|\, \cdot & \distas & \IG\left( 1.5, 1 +
  \delta_{jl}^2 \right),\\
  \tau^2_{\epsilon} \,|\, \cdot & \distas &
  \IG\left\{ 1 + \frac{1}{2} \sum_{i=1}^I\sum_{j=1}^J n_{ij}, ~
  10^{-5} + \frac{1}{2} \sum_{i=1}^I\sum_{j=1}^J\sum_{k=1}^{n_{ij}}
  \left(\bld{Z}_i  \left[ 
    \begin{array}{cc} 
    \bs{\phi}\\
    \bs{\theta}_j 
  \end{array}
  \right] - \psi_{ijkl} \right)^2 \right\}.
\end{eqnarray}
Hence, the variance parameters can be drawn independently.

\bibliographystyle{mrm}
\bibliography{badnews,mripapers,gsk,cic,climate}

\newpage

\begin{table}[p]
  \caption{Median $\ktrans$ values from the standard analysis (R =
  responder, NR = non-responder).}
  \begin{center}
    \begin{tabular}{c|cccccccccccc}
      \hline\hline
      Patient ID & 1 & 2 & 3 & 4 & 5 & 6 & 7 & 8 & 9 & 10 & 11 & 12 \\
      \hline
      pathological & R & R & R & NR & NR & R & NR & NR & R & NR & NR & R \\ 
      \hline
      pre & 0.208 & 0.355 & 0.255 & 0.230 & 0.199 & 0.154 & 0.264 & 0.198 & 
      0.305 & 0.267 & 0.432 & 0.174 \\
      post & 0.161 & 0.120 & 0.031 & 0.245 & 0.208 & 0.173 & 0.327 & 0.223 & 
      0.122 & 0.221 & 0.111 & 0.113 \\
      \hline\hline
    \end{tabular}
  \end{center}
  \label{tab:standard-analysis}
\end{table}

\newpage

\section*{Figure Captions}

\begin{enumerate}

\item Schematic overview of the Bayesian hierarchical model for the
  observed contrast agent concentration time curves.

\item Contrast concentration time curves (CTCs) for pre- and
  post-treatment scans in three different subjects.  Solid lines
  indicate the voxel-specific fit to the CTC, dashed lines the
  combined patient- and treatment-specific CTCs, and dotted lines the
  global pre- and post-treatment median CTCs for the entire study.

\item Samples from the marginal posterior distributions of $\ktrans$
  at the study level.  At pre-treatment $\ktrans$ is given by
  $\exp(\alpha_1)$ and at post-treatment $\ktrans$ is given by
  $\exp(\alpha_1+\beta_1)$.

\item Samples from the marginal posterior distributions of $\ktrans$
  at the patient level.  At pre-treatment $\ktrans$ is given by
  $\exp(\alpha_1+\gamma_{j1})$ for patient~$j$ and at post-treatment
  $\ktrans$ is given by
  $\exp(\alpha_1+\beta_1+\gamma_{j1}+\delta_{j1})$ for patient~$j$.

\item Smoothed histograms summarizing the values of the posterior
  median $\ktrans$ at the voxel level.  At pre-treatment $\ktrans$ is
  given by $\exp(\alpha_1+\gamma_{j1}+\epsilon_{1jk1})$ for scan~$1$,
  patient~$j$ and voxel~$k$.  At post-treatment $\ktrans$ is given by
  $\exp(\alpha_1+\beta_1+\gamma_{1j}+\delta_{1j}+\epsilon_{12jk})$ for
  scan~$2$, patient~$j$ and voxel~$k$.  The $x$-axis has been
  restricted to $[0,1]$ for visualization.

\item Smoothed histograms summarizing the values of $\ktrans$ from
  voxel-wise non-linear regression analysis.  The $x$-axis has been
  restricted to $[0,1]$ for visualization.

\end{enumerate}

\newpage

\end{document}